# Modeling Warfare in Social Animals: A "Chemical" Approach

Alisa Santarlasci (a,c), Gianluca Martelloni (b,c), Filippo Frizzi (d), Giacomo Santini (d), Franco Bagnoli (b,c)

   a) Dept. of Information Engineering, University of Florence, via S. Marta 1, 50139 Firenze, Italy.
   b) Dept. of Physics and Astronomy, University of Florence, via G. Sansone 1, 50019 Sesto Fiorentino and INFN, sez. Firenze.
   c) Center for the Study of Complex Dynamics, University of Florence, via G. Sansone 1, 50019 Sesto Fiorentino.
   d) Dept. of Biology, University of Florence, via Madonna del Piano 6, Sesto Fiorentino.

## Abstract

The aim of our study is to describe the dynamics of ant battles, with reference to laboratory experiments, by means of a chemical stochastic model. We focus on ants behavior as an interesting topic for their aptitude to propagate easily to new habitats. In order to predict the ecological evolution of invasive species and their relative fast spreading, a description of their successful strategies, also considering their competition with other ant species is necessary. In our work we want to describe the interactions between two groups of different ant species, with different war strategies, as observed in our experiments. The proposed chemical model considers the single ant individuals and fighting groups in a way similar to atoms and molecules, respectively, considering that ant fighting groups remain stable for a relative long time. Starting from a system of differential non-linear equations (DE), derived from the chemical reactions, we obtain a mean field description of the system. This deterministic approach is valid when the number of individuals of each species is large in the considered unit, while in our experiments we consider battles of at most 10 vs. 10 individuals, due to the difficulties in following the individual behavior in a large assembly. Therefore, we also adapt a Gillespie algorithm to reproduce the fluctuations around the mean field description. The DE schematization is exploited to characterize the stochastic model. The set of reaction constants of chemical equations, obtained by means of a minimization algorithm between the DE and the experimental data, are used by the Gillespie algorithm to generate the stochastic trajectories. We then fit the stochastic paths with the DE, in order to analyze the variability of the parameters and therefore their variance. Finally, we estimate the goodness of the applied methodology and we confirm that the stochastic approach must be considered for a correct description of the observed ant fighting dynamics. With respect to other war models (e.g., Lanchester's ones), our chemical model considers all phases of the battle and not only casualties. Therefore, we can count on more experimental data, but we also have more parameters to fit. In any case, our model allows a much more detailed description of the fights.

# 1. Introduction

Ants are an interesting topic as the primer of a possible significant ecological transformation and for the extraordinary strategies used to achieve their ecological success. The latter is also linked to the capacity to overcome other ant species. Actually, the warfare art of ants is usually known for the copious strategies used and the impressive violence exhibited.

In particular, an invasive ant species is characterized by its ability to shape enormous colonies that can expand in a supercolony of a number of interconnected nests. Thus, the ants become ecologically dominants through an exaggerate cooperative behavior to colonize new territories and to attack their enemies.

In order to predict the ecological evolution of invasive species and their relative fast spreading, a description of their successful strategies, also considering their competition with other ant species, is necessary. Furthermore, detailed mathematical models of ant fighting dynamics may also benefit from the understanding of ecological interactions and community behavior. Our studies are focused on two species that share the same habitat but with different fighting behavior and strategies: the invasive ant *Lasius Neglectus* and *Lasius Paralienus.*

Asia Minor is the most likely native environment of *Lasius Neglectus* (Seifert, 2000) where it co-occurs with his relative *Lasius Turcicus*: in few decades it has spread all over Europe from Spain to the northern zones where it can survive at frost temperature (-5 °C) (Ugelvig *et al*., 2008). Moreover, in consequence of a low number of jump dispersal due to human transportation, this species infests urban disturbed habitats where it is eradicating most native ants and other insect populations, changing the ecosystem equilibrium and damaging trees and culture farms (Cremer *et al*., 2008).

*L. Neglectus* is a formidable machine due to the social structure of its supercolony, where queens mate in the nest and disperse on foot accompanied by workers (colony budding). In this way colonies occupy multiple interconnected nests with many queens (polygyny) (Cremer *et al*., 2008, Tsutsui and Suarez, 2002). By mating within the native colony, the invasive ant reproduction is independent from the presence of other colonies and consequently queens don't need functional wings and a large bodies with fat reserves to survive in isolate lands. Free movement among the nests facilitates the free mixing of the individuals and therefore they are genetically unrelated as randomly sampled individuals (Cremer *et al*., 2008).

*L. Neglectus* perfectly exhibits the "invasive ant syndrome" with a cooperative behavior and a total absence of aggression among ants from different nests. To explain this behavior feature, the reduction of genetic variation at loci coding for cuticular hydrocarbons (their chemical identification signal) was supposed and then confirmed, along with the abundance of low volatile long–chain hydrocarbons, less informative as recognition cues (Cremer *et al*., 2008).

Even if the story of *L. Neglectus* is quite recent, his pre-adaptation before-invasion traits were sufficient to forecast his diffusion all over Europe. Few events of unaware human transport, his selected ability to adapt to human environment, a high range of temperature sustained, the absence of inter-nest aggression and a strong and organized aggression towards other ant species, support his ecological success.

*L. Paralienus* (*Seifert,* 1992) is an endemic species of central Europe, its distributional center is probably the Balkans and it expands in Turkey too. It predominates continental climate and it is absent from urban areas in Central Europe. *L. Paralienus* is a monogynous species and it shows the biggest queens of all Palaearctic species. It forms colonies not very wide and usually it exhibits a non-dominant behavior.

In general, social animals are often involved in group combats, where members of the same group can cooperate during fight against conspecific or heterospecific opponents. Fighting in groups determines a fundamental change in the dynamics of the battle, because the outcome of the fight relies both on individual fighting ability and on group sizes. In the simplest case, a battle is the result of a series of individual duels, while in case of more organized armies, the individuals can cooperate in facing a single opponent. In the latter circumstance it is evident that group size may have a disproportionate importance over individual fighting ability.

To describe the battles between two armies, Lanchester (1916) proposed a simple mathematical model in two versions denominated "linear law", Eq. (1), and "square law", Eq (2). Given two groups of opponents, whose number of individuals is $m$ and $n$, the death rates $dm/dt$ and $dn/dt$ according to the linear and the square law can be described as:

$$\begin{cases} dm/dt = -\alpha_n \cdot n \cdot m \\ dn/dt = -\alpha_m \cdot m \cdot n \end{cases} \Rightarrow \quad \alpha_m (m - m_0) = \alpha_n (n - n_0), \qquad (1)$$

$$\begin{cases} dm/dt = -\alpha_n \cdot n \\ dn/dt = -\alpha_m \cdot m \end{cases} \Rightarrow \quad \alpha_m (m^2 - m_0^2) = \alpha_n (n^2 - n_0^2), \qquad (2)$$

where $\alpha_n$ and $\alpha_m$ are the fighting abilities of specimens of the $n$ and $m$ groups, respectively, while $m_0$ and $n_0$ are the initial conditions. Under the linear law, Eq. (1), it is implicitly assumed that the battle is represented by a series of individual duels, so that outnumbering individuals of the more numerous group remain unengaged until an opponent becomes available. On the contrary, the square law, Eq. (2), assumes that members of the more numerous group gang together against individual opponents.

Lanchester models have been extensively used to describe battle outcomes or plan warfare tactics (Bracken, 1995; Fricker, 1998, Johnson and MacKay, 2008). Starting from the work of Franks and Patridge (1993), who first unveiled their usefulness in behavioral ecology, Lanchester theory has been also applied to animal contest. Not surprisingly, a relevant number of studies addressing the application of Lanchester theories to animal conflicts contemplate ants (Barchelor and Briffa, 2010; Frank and Partridge, 1993; McGlynn, 2000; Plowes and Adams, 2005; Wilson *et al.,* 2001; Withehouse and Jaffe, 1996;).

Lanchester models are applied for the description of intraspecific fight (Plowes and Adams, 2005; Batchelor and Briffa, 2010, 2011; Batchelor *et al*., 2012) as well as to fit interspecific interactions (Franks and Partridge., 1993). Both the linear and the square Lanchester laws are applied to ant contests, proving in line with the experimental data (Franks and Partridge, 1993; Withehouse and Jaffe, 1996).

In none of these studies, however, attempts to fit directly the observed dynamics of a battle with theoretical expectations have been made, rather evidences in favor of one strategy or the other were inferred from indirect observations, *e.g.,* the size and number of individuals involved.

One of the strengths of Lanchester models is their simplicity which make them basic, although reasonably realistic benchmark for opposing battle types. The Lanchester model shows a rough description of the ant fighting and doesn't take in to account the specific dynamic of the specimens during the battle. With our approach we shall describe a more realistic ant behavior with their capacity to cooperate, casting a glance inside the battlefield.

Our studies start from some experiments in which we arranged a fighting match, in a petri dish with a diameter of 10 cm, between *L. Neglectus* and *L. Paralienus* using 5 up to a maximum of 15 individuals for each species. In order to observe the dynamics of the system, *i.e.*, to follow the ant movements and the group formation, we focus on 10 vs. 10 battles that represent a balance between the size of the system and its dynamical observability. To describe the dynamics inside the system and the formation of fighting groups with their time trend, we start from a set of chemical equations which encode the interactions among individual entities, *i.e.*, we consider the isolated ant individuals and the groups formed during the battle as chemical species. For example, two comrades can stick to an enemy ant and so we consider them as a chemical species (molecule) composed by three ants. From the chemical reactions we deduce a system of differential non-linear equations (DE) that allows a mean field description of the system, *i.e.*, the evolution of the considered species over time. Actually, this deterministic approach is valid when the state variables, in our case the number of individuals of each species, are sufficiently large (Campillo and Lobry., 2012). In our experiments the number of ants is small, but the schematization with differential equations is then exploited to characterize the stochastic model that we introduce for modeling also the fluctuations of this system. For this goal, we adapt the Gillespie algorithm (Gillespie, *1977*). In order to do that, we need a set of parameters, *i.e.,* the reaction constants of the chemical equations, with which we can generate an ensemble of trajectories with the Gillespie algorithm. To identify the parameters, starting from DE, we use the Flexible Simplex optimization algorithm (Marsili-Libelli, 1992). Then we fit the stochastic paths resulted from the Gillespie algorithm with the DE, in order to analyze the parameters variability and therefore the variance. With this method we estimate the goodness of the applied methodology and we confirm that the stochastic approach must be considered for a correct characterization of the observed ant battle dynamics.

The paper is organized as follows: in the following Sections 2.1. and 2.2 we introduce the experimental set-up and the chemical model discussing its mathematical formulation. In section 3.the stochastic approach is reported. In Section 4. we compare the deterministic simulations with the stochastic outputs. Finally, in section 5., we discuss the results and report our conclusions.

## 2. Materials and Methods

2.1. Experimental model: Ant sampling and site

To perform our experiments we collected ants during July/ August 2012 in Prato (Northern Tuscany, Italy, 43° 52' 46"N, 11° 05' 50"E), where abundant populations of *L. Neglectus* and *L. Paralienus* colonize urban garden trees. The collection occurred during the first warm morning hours. Each species was collected from a single large nest at 200mt distance one to the other. All the two species are monomorphic with reduced intraspecific differences in the size of ants. Then they were stored in a falcon tube in groups of 10 specimens with the relief of wet cotton ball until the experiments were performed. Previously the petri was cleaned with alcohol to remove impurities and the walls were coated with Fluon. We releases the two samples of 10 *L. Neglectus* and 10 *L. Paralienus* in a 10 cm petri dish. Then, we recorded the fighting with a fixed camera for an hour in all.

The video were analyzed sampling over time the number of individuals of each defined species involved in each fighting. Let us denote L. *Paralienus* with *A* and L. *Neglectus* with *B*. The groups *AB*, *ABB* and *ABBB* denote the "chemical species" in which one *A* fights with one *B*, one *A* with two *B*'s and one *A* with three *B*'s, respectively. We also noted the changes of the groups in consequence of escape or mortality (rather rare in our observations). We measured the lifetime of the fighting groups taking in to account those with a time-lapse longer than 20 seconds (Table 1). The examination was done visually, recognizing each species from their dimensions and color: *L. Neglectus* is slightly yellow and smaller than *L. Paralienus*. The two species exhibit a different behavior: in most of cases *L. Neglectus* is the first aggressor and cooperates, but it also has the greater mortality. *L. Paralienus* has the aptitude to escape and avoid the fighting, it doesn't cooperate but it is bigger and quite stronger than *L. Neglectus*. Actually, *L. Neglectus* is smaller compared to most ant species and his success depends on its high densities of population and aggressiveness employed in the combats. Especially the ability to cooperate by two, tree and also four ants, allow *L. Neglectus* to prevail in the fighting as it hang on to the enemy biting legs or antenna.

| Seconds | A | B | AB | ABB | ABBB |
|---|---|---|---|---|---|
| 577 | 5 | 1 | 1 | 4 | 0 |
| 640 | 5 | 3 | 3 | 2 | 0 |
| 660 | 5 | 5 | 5 | 0 | 0 |
| 750 | 5 | 3 | 4 | 0 | 1 |
| 840 | 5 | 2 | 3 | 1 | 1 |
| 855 | 6 | 3 | 2 | 1 | 1 |
| 890 | 6 | 2 | 2 | 0 | 2 |
| 960 | 6 | 2 | 2 | 0 | 2 |
| 990 | 5 | 3 | 3 | 2 | 0 |
| 1020 | 5 | 3 | 3 | 2 | 0 |
| 1050 | 4 | 2 | 4 | 2 | 0 |
| 1080 | 4 | 2 | 4 | 2 | 0 |
| 1110 | 5 | 1 | 3 | 2 | 0 |
| 1140 | 5 | 0 | 3 | 2 | 0 |
| 1170 | 5 | 0 | 3 | 2 | 0 |
| 1200 | 7 | 2 | 1 | 2 | 0 |

**Table 1.** Extracting data from an experiment: in the first column the time (in seconds) when a reaction occur, then the total number of individuals of the species A and B and of the forming groups AB, ABB, ABBB.

## 2.2. Mathematical model: the deterministic model

In order to study the dynamics of the ants fighting we propose a chemical model, which explicitly incorporates the interactions among individuals. During the fighting, ants arrange in steady groups, that often remain stable for a relative long time like chemical compounds. The aim of the analysis is to describe, within a minimalist self-consistent dynamical equation, what happen when two groups of different ants species, with different strategies of attack, interact, as seen in our experiments.

The model consists in a collection of chemical equations, which encode for the interactions among individual entities. They can be translate in a set of ordinary differential equations considering the law of mass action. The first reaction happens when an individual of *L. Neglectus* (chemical species B) establishes a strong tie in fight with an individual of *L. Paralienus* (chemical species A) to form a new group or chemical species AB. The latter behavior can be expressed as

$$A + B \underset{k_2}{\overset{k_1}{\leftrightarrow}} AB, \tag{3}$$

where $k_1$ and $k_2$ are respectively the reaction constants of the direct and reverse reactions (we omit to explicitly illustrate the reactions constants in the following). The outcome of a duel can lead to the death of A,

$$AB \overset{k_3}{\rightarrow} B, \tag{4}$$

or the death of B,

$$AB \overset{k_4}{\rightarrow} A. \tag{5}$$

Once a group AB is formed, a second B can participate to the fight. We describe the appearance of a group ABB by means of a reversible reaction

$$AB + B \underset{k_6}{\overset{k_5}{\leftrightarrow}} ABB, \tag{6}$$

since in the experiments we can also observe a B that detaches from the group.

In a fighting group ABB an item B can die, so we describe this scenario with an irreversible reaction

$$ABB \overset{k_7}{\rightarrow} AB. \tag{7}$$

Then we add the possibility, as we observe in the experiments, that also an ant A dies in consequence of a fighting with two ants B. In this case the group ABB dissolves,

$$ABB \overset{k_8}{\rightarrow} 2B. \tag{8}$$

Another recurring possibility from a state ABB is the dissolution of the group. Furthermore, we can observe the opposite reaction, *i.e.*, two B attack an A, not actually simultaneously but in a very short interval with respect to the observation time in the experiments. Therefore we add the reaction

$$ABB \underset{k_{10}}{\overset{k_9}{\leftrightarrow}} A + 2B, \tag{9}$$

The observations also show the sticking of three B's with an A, as described in the following reversible reaction, in which it is also possible that a B escapes from the group,

$$ABB + B \underset{k_{12}}{\overset{k_{11}}{\leftrightarrow}} ABBB. \tag{10}$$

An individual A can die as consequence of the fighting with three B's and the group dissolves, *i.e.*,

$$ABBB \overset{k_{13}}{\rightarrow} 3B. \tag{11}$$

Finally another possibility is the detachment of two B's from the group. Considering also the opposite reaction, we have:

$$ABBB \underset{k_{15}}{\overset{k_{14}}{\leftrightarrow}} AB + 2B, \tag{12}$$

where it is assumed that the two B's attach to the group AB separately, but in a very short time compared to the reaction observed timing.

The above chemical reactions can be translated into a set of non-linear ordinary differential equations.

Let $x = A$, $y = B$, $z = AB$, $u = ABB$ and $v = ABBB$. The differential equation system is

$$\dot{x} = -k_1 \cdot x \cdot y - k_{10} \cdot x \cdot y^2 + (k_2 + k_4) \cdot z + k_9 \cdot u, \tag{13}$$

$$\begin{aligned}\dot{y} = &-k_1 \cdot x \cdot y - 2 \cdot k_{10} \cdot x \cdot y^2 + (k_2 + k_3) \cdot z - k_5 \cdot y \cdot z - 2 \cdot k_{15} \cdot y^2 \cdot z + \\ &+ (k_6 + 2 \cdot k_8 + 2 \cdot k_9) \cdot u - k_{11} \cdot y \cdot u + (k_{12} + 3 \cdot k_{13} + 2 \cdot k_{14}) \cdot v\end{aligned}, \tag{14}$$

$$\dot{z} = k_1 \cdot x \cdot y - k_5 \cdot y \cdot z - (k_2 + k_3 + k_4) \cdot z + (k_6 + k_7) \cdot u - k_{15} \cdot y^2 \cdot z + k_{14} \cdot v, \tag{15}$$

$$\dot{u} = k_{10} \cdot x \cdot y^2 + k_5 \cdot y \cdot z - k_{11} \cdot y \cdot u - (k_6 + k_7 + k_8 + k_9) \cdot u + k_{12} \cdot v, \tag{16}$$

$$\dot{v} = k_{11} \cdot y \cdot u - (k_{12} + k_{13} + k_{14}) \cdot v + k_{15} \cdot y^2 \cdot z, \tag{17}$$

where

$$\dot{x} = \frac{dx}{dt}, \quad \dot{y} = \frac{dy}{dt}, \quad \dot{z} = \frac{dz}{dt}, \quad \dot{u} = \frac{du}{dt}, \quad \dot{v} = \frac{dv}{dt},$$

represent the time derivatives, *i.e.*, the growth rate for each considered chemical species.

In order to integrate this ordinary differential system we use the numerical method of Cash and Karp (Cash and Karp, 1990) checking that the fourth and fifth order solutions provide the same results. This method is a member of the Runge-Kutta family of ordinary differential equation solvers.

The system, represented by Eqs. (13-17), does not exhibit fixed points, *i.e.*, the stationary solution ($dx/dt = 0$, $dy/dt = 0$, $dz/dt = 0$, $du/dt = 0$, $dv/dt = 0$) depends on the initial conditions (see Figure 11(a,b,c,d,e). This dependence on initial condition is consistent with the biological point of view, *i.e.*, with our experimental observations repeating the battles after varying the number of individuals. The number of survivors of the winning species depends on the initial number of individuals involved in the battle.

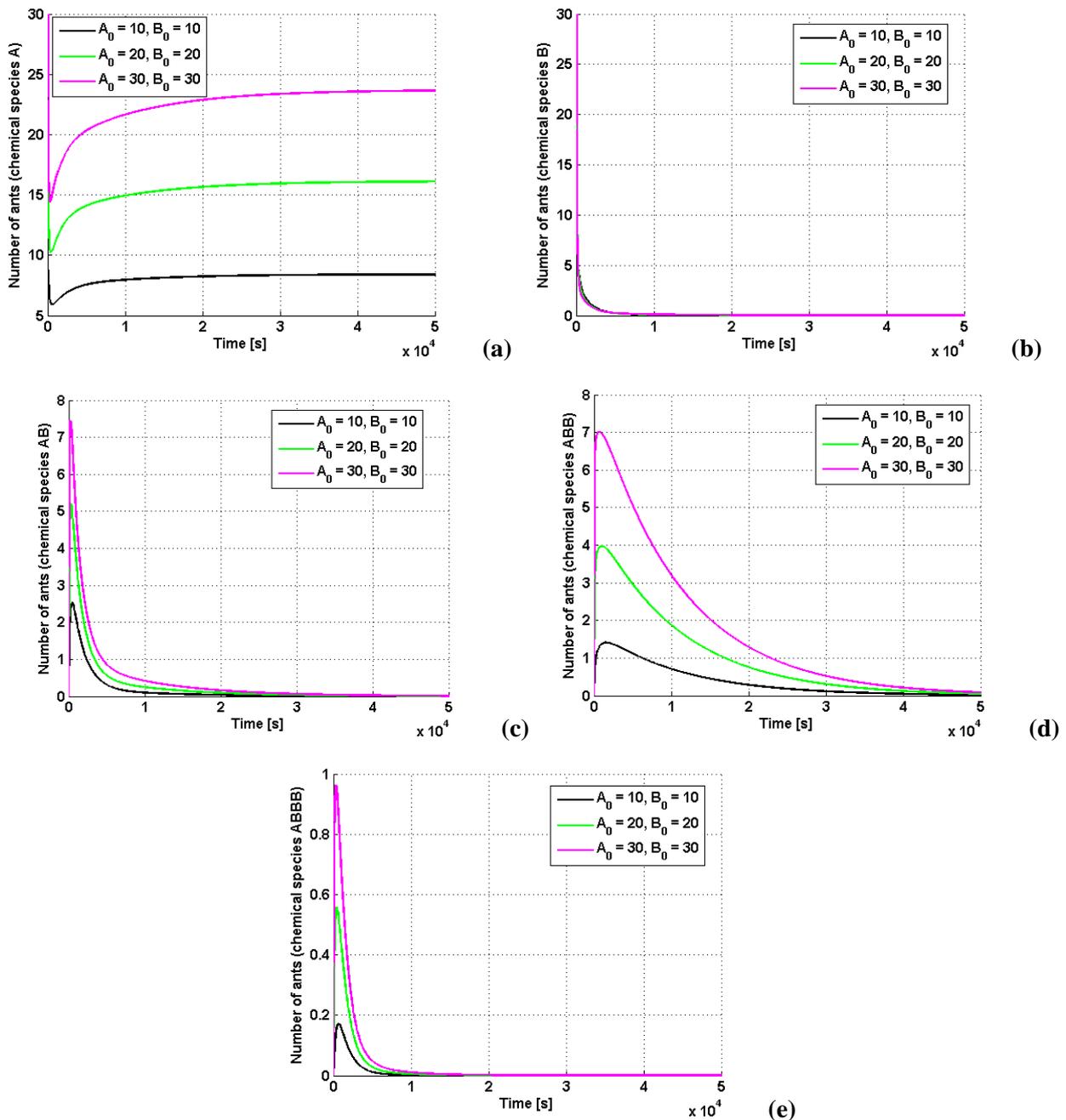

**Fig. 1.** In a, b, c, d, e the chemical species respectively of species A, B, AB, ABB, ABBB over the time up to the steady state for three value of the initial conditions.

We proceed with the model identification to obtain the best parameters of the system, minimizing the error between the experimental "chemical" population data and the model output. For this purpose we use a parametric optimization algorithm (Marsili-Libelli, 1992), the Simplex Flexible Algorithm (SFA). The error functional $F(\mathbf{P})$, that are minimized by SFA varying the parameter $\mathbf{P} = [k_1, \ldots, k_j, \ldots, k_{15}]$, is defined as

$$arg\left[\min_{\mathbf{P}} F(\mathbf{P})\right] = arg\left[\min_{\mathbf{P}} \sum_{si=1}^{5} \frac{1}{N} \sum_{i=1}^{N} \left(x_{si,i}^{\exp} - x_{si,i}^{\mod}(\mathbf{P})\right)^2\right] \quad \text{with} \quad (x_1, x_2, x_3, x_4, x_5) = (x, y, z, u, v) \quad (18)$$

where $x^{\exp}$ and $x^{\mod}$ indicate the experimental and the model values, respectively, for each species $si$ and for each population value $i = 1, 2, \ldots, N$.

## 3. Mathematical model: the stochastic approach

The deterministic approach is not sufficient to completely characterize the ant battle dynamics since, in our experiments, we observe fluctuations in terms of ant groups that form over time. We report two experiments of 10 vs. 10 ants (see Figure 2,3) and some stochastic simulations (described in the following) in order to show the fluctuations of the system (see Figures 4,5). It is therefore necessary to take into account a stochastic approach.

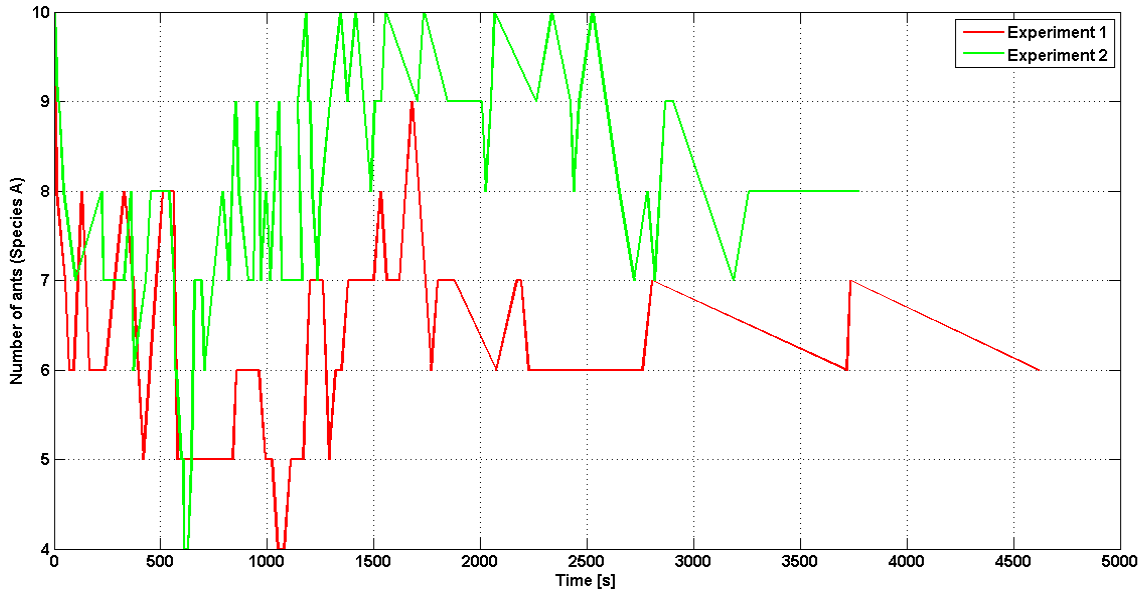

**Fig. 2.** The chemical species A over the time obtained from two experiments.

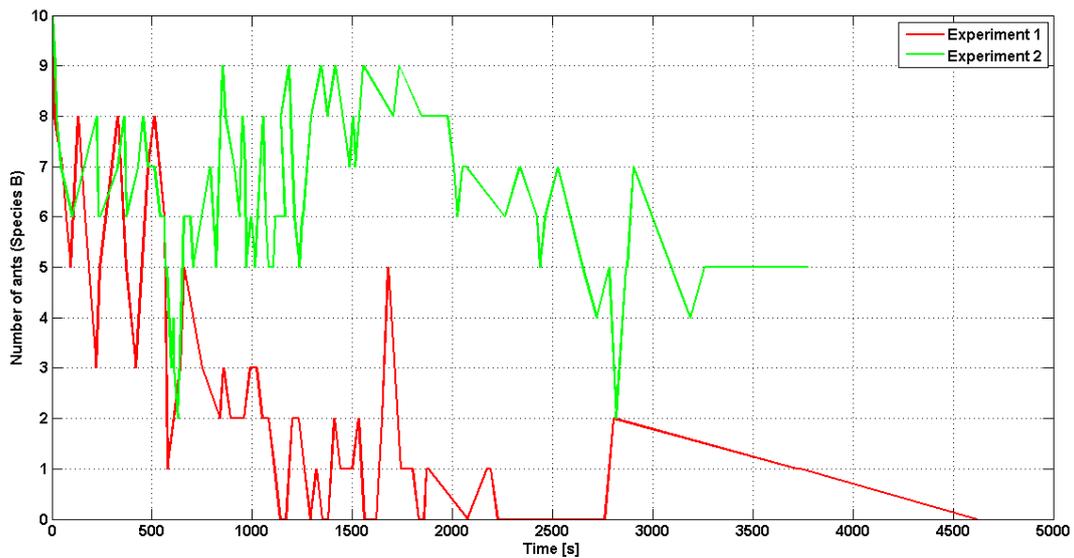

**Fig. 3.** The chemical species B over the time obtained from two experiments.

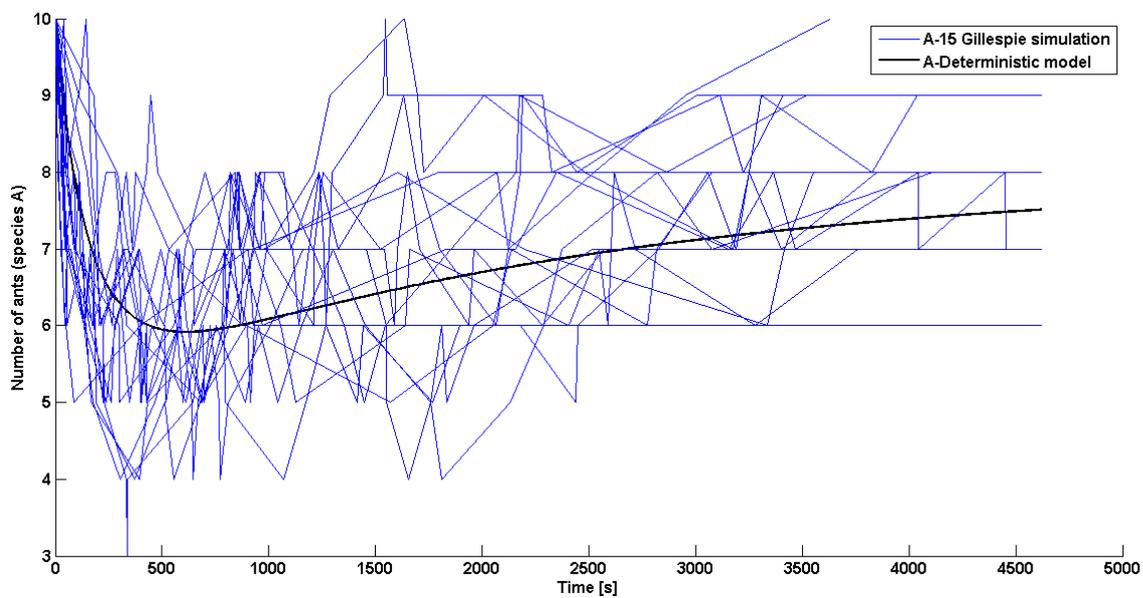

**Fig. 4.** The chemical species A over the time obtained with the deterministic model (mean field) and the fluctuations generated with the Gillespie algorithm (15 simulations).

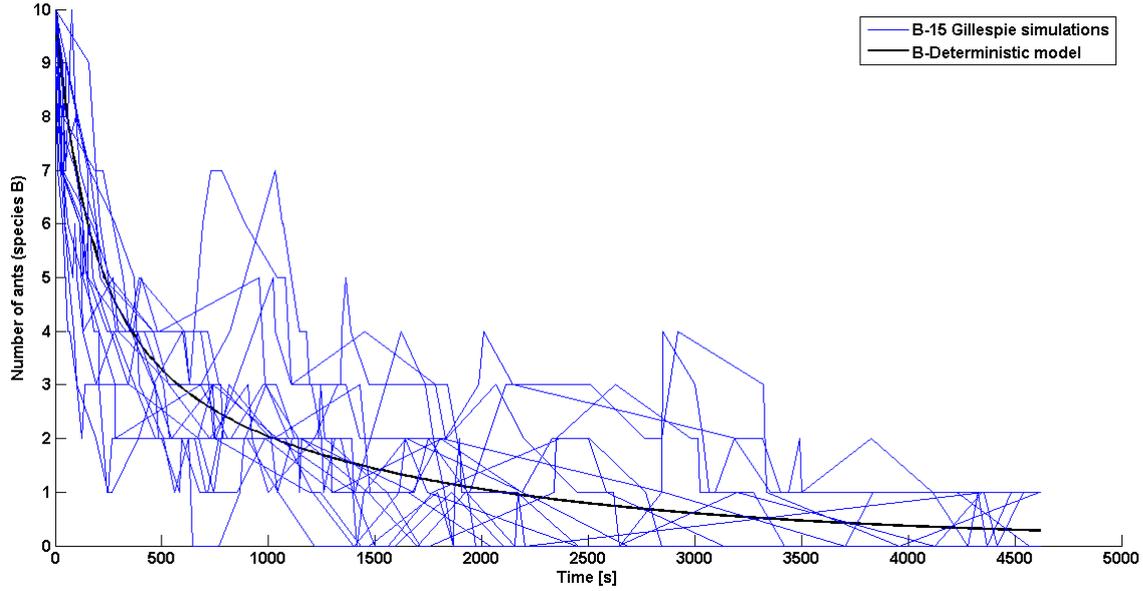

**Fig. 5.** The chemical species B over the time obtained with the deterministic model (mean field) and the fluctuations generated with the Gillespie algorithm (15 simulations).

For this purpose we use the Gillespie's direct method (Gillespie 1977), *i.e.*, an algorithm that generates a trajectory, statistically correct, of the master equations. The latter represent a set of differential equations that take into account the variation of the probabilities, that the system occupies each one of a discrete set of states, over time that is considered a continuous variable. This method explicitly simulates each reaction, giving a stochastic formulation of chemical kinetics based on the theory of collisions. In our cases, giving the species *si* and the 15 reactions expressed by means of the chemical Eqs. (3-12) with reaction constants $k_j$, the propensity function is defined as $F_j = k_j \cdot H_j$, where $H_j$ is the number of distinct individual (of each specie) reactant combinations. Then, considering a time interval $\Delta t$, the reaction probability density function is defined as:

$$\begin{cases} P(\Delta t, j) = P_1(\Delta t) \cdot P_2(j) \\ P_1(\Delta t) = F_0 \cdot \exp(-F_0 \cdot \Delta t) \\ P_2(j) = \dfrac{F_j}{F_0} \\ F_0 = \sum_{j=1}^{15} F_j \end{cases} \qquad . \tag{19}$$

Noting that the reaction probability density function is separable in two parts, i.e., an exponential distribution of time reactions ($P_1$) and the normalized propensity function ($P_2$), the Gillespie algorithm can be implemented according two steps by choosing $\Delta t$ and $j$ (i.e., the reaction) as follows:

$$\Delta t = \frac{1}{F_0} \cdot \ln\left(\frac{1}{r_1}\right) \tag{20}$$

and $j$ as the integer for which

$$\sum_1^{j-1} F_j < r_2 \cdot F_0 \leq \sum_1^{j} F_j, \tag{21}$$

where $r_1$ and $r_2$ are uniform random number between 0 and 1. Then, in our cases, the propensity vector of reactions is:

$$\begin{bmatrix} k_1 \cdot A \cdot B \\ k_2 \cdot AB \\ k_3 \cdot AB \\ k_4 \cdot AB \\ k_5 \cdot B \cdot AB \\ k_6 \cdot ABB \\ k_7 \cdot ABB \\ k_8 \cdot ABB \\ k_9 \cdot ABB \\ k_{10} \cdot A \cdot B \cdot (B-1)/2 \\ k_{11} \cdot B \cdot ABB \\ k_{12} \cdot ABBB \\ k_{13} \cdot ABBB \\ k_{14} \cdot ABBB \\ k_{15} \cdot AB \cdot B \cdot (B-1)/2 \end{bmatrix} \tag{22}$$

Thus we adapt the Gillespie algorithm considering the optimized parameters (reaction constants) obtained with the SFA. We generated 100 trajectories for each considered chemical species in order to fit each one with the deterministic model using the SFA. In this way we obtained 100 values for each reaction constant $k_j$, used to perform the statistical analysis and to test the goodness of the proposed method, see Figure 6. In other words we exploit this procedure to compare the deterministic model with the stochastic one and to test the reliability of the reaction constants, obtained with SFA.

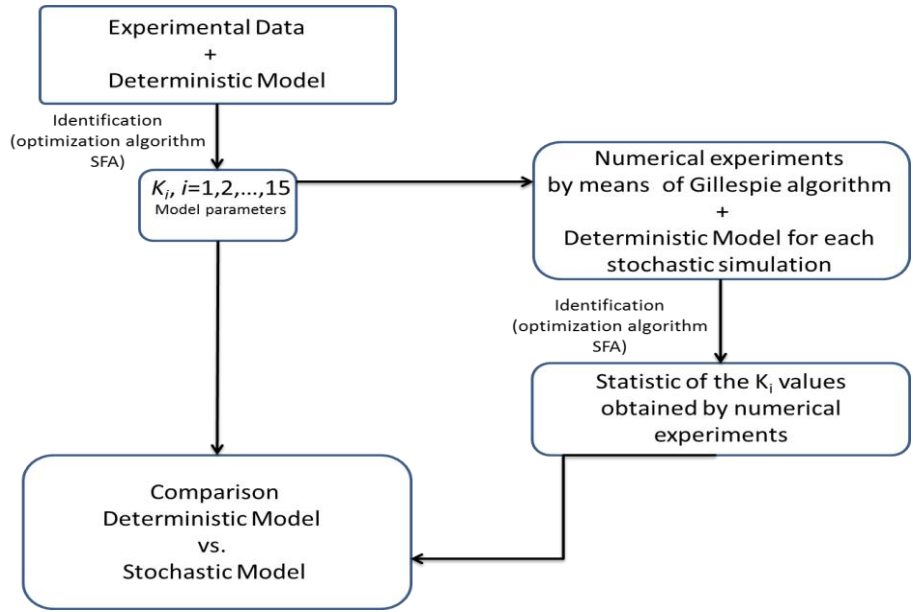

**Fig. 6.** The scheme of the proposed methodology in order to compare the chemical model, expressed by means of a system of nonlinear differential equations, and the stochastic one.

## 4. Simulation results and tuning

The optimized parameters, obtained by means of SFA, are reported in Table 2 (column one), and we show, in Figures 7-11, the number of individuals over time, deduced from the experimental results, fitted by the optimized solutions of the deterministic model.

| $k_i$ | True values by parametric identification of DE | Mean of the parameters by stochastic simulation | R-square (Gaussian fit) |
|---|---|---|---|
| $k_1$ | 0.0002438 | 0.0002644 | 0.9601 |
| $k_2$ | 0.0006932 | 0.0007122 | 0.9986 |
| $k_3$ | 7.7388e-05 | 7.211e-005 | 0.976 |
| $k_4$ | 0.001211 | 0.001243 | 0.9728 |
| $k_5$ | 6.94257e-06 | 6.69e-006 | 0.9408 |
| $k_6$ | 4.75385e-06 | 4.6e-006 | 0.9813 |
| $k_7$ | 4.50353e-05 | 4.06e-005 | 0.9458 |
| $k_8$ | 3.40730e-05 | 2.974e-005 | 0.9669 |
| $k_9$ | 6.38355e-06 | 6.274e-006 | 0.9583 |
| $k_{10}$ | 1.05249e-05 | 9.82e-006 | 0.9213 |
| $k_{11}$ | 3.13041e-05 | 3.582e-005 | 0.995 |
| $k_{12}$ | 0.00040597 | 0.0004277 | 0.9647 |
| $k_{13}$ | 0.0009040 | 0.0008634 | 0.9426 |
| $k_{14}$ | 1.22548e-06 | 1.191e-006 | 0.9279 |
| $k_{15}$ | 6.53508e-06 | 6.197e-006 | 0.7305 |

**Table 2.** In the first column the value of the optimized reaction constants, in the second one the mean of $k_i$ obtained by stochastic view and finally the R-square by Gaussian fit of the distribution of $k_i$.

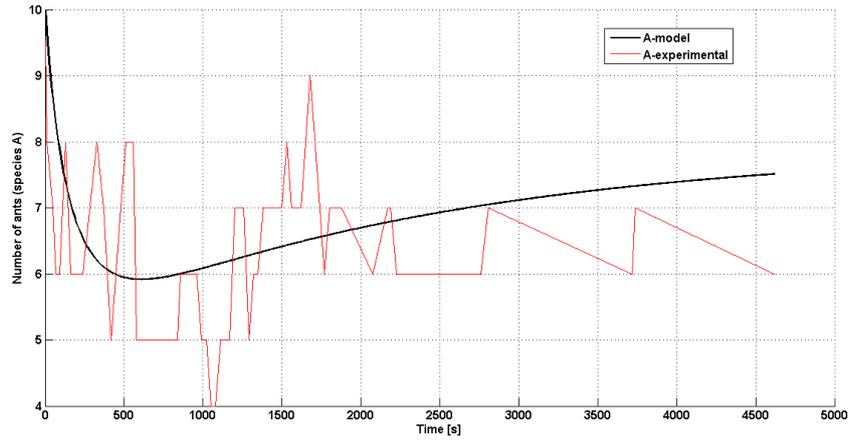

**Fig. 7.** The experimental data vs. the solution of the deterministic model identify by means of SFA for the chemical species A.

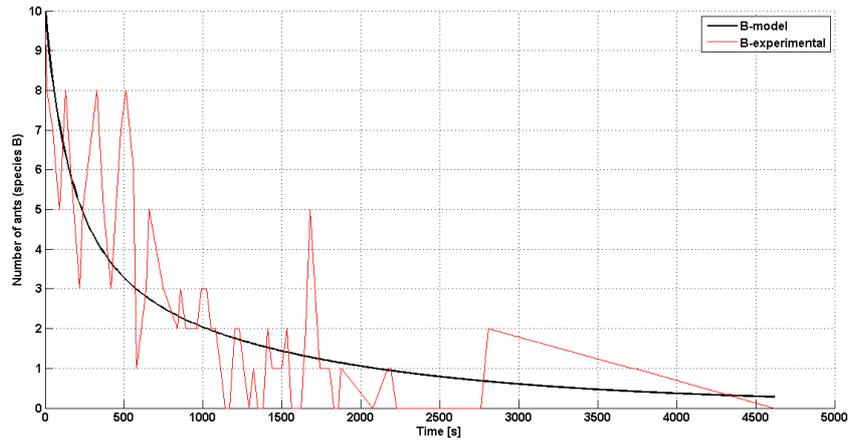

**Fig. 8.** The experimental data vs. the solution of the deterministic model identify by means of SFA for the chemical species B.

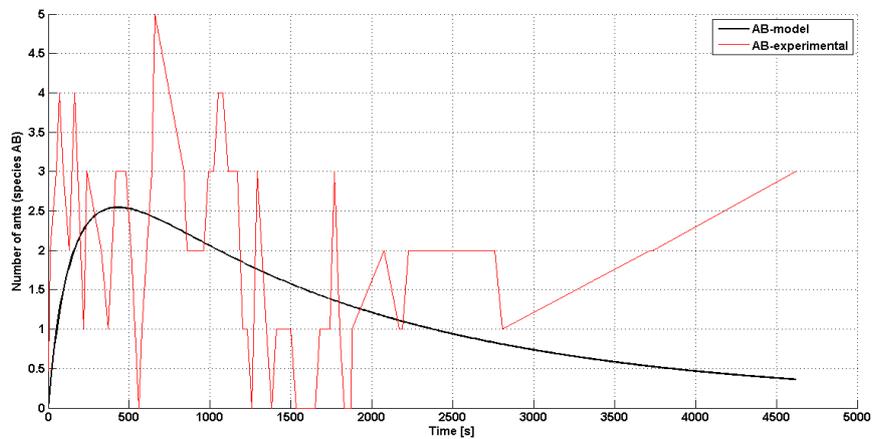

**Fig. 9.** The experimental data vs. the solution of the deterministic model identify by means of SFA for the chemical species AB.

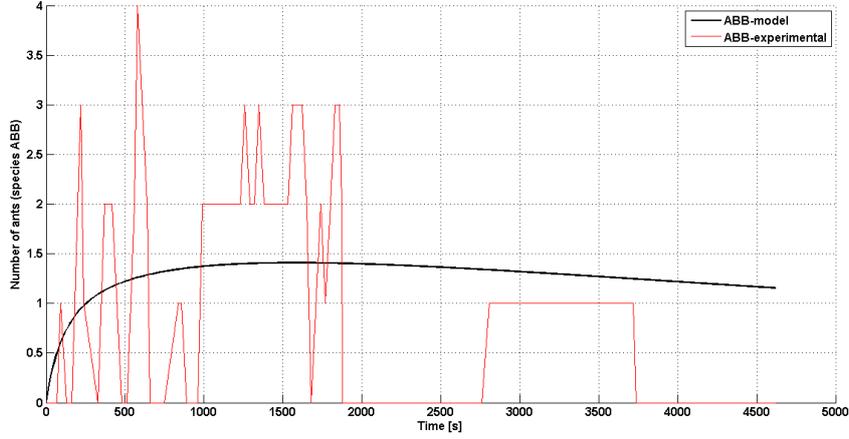

**Fig. 10.** The experimental data vs. the solution of the deterministic model identify by means of SFA for the chemical species ABB.

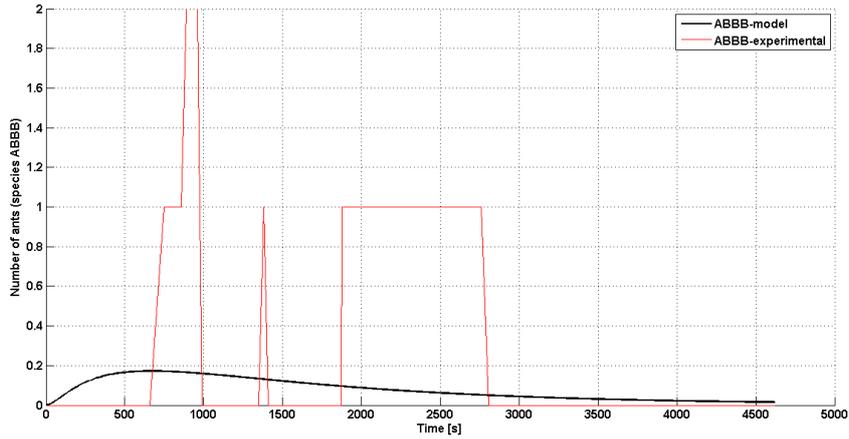

**Fig. 11.** The experimental data vs. the solution of the deterministic model identify by means of SFA for the chemical species ABBB.

To verify the goodness of our model we achieve a comparison with a modified Lanchester model that takes into account Eqs. (1) and (2), *i.e.*,

$$\begin{cases} dA^*/dt = -k_{l1} \cdot B^* - k_{l2} \cdot A^* \cdot B^* \\ dB^*/dt = -k_{l3} \cdot A^* - k_{l4} \cdot A^* \cdot B^* \end{cases} \qquad (23)$$

Considering the ratio between the two derivatives of Eq. (23) and integrating in $A$ and $B$, we obtain

$$\frac{(A^* - A_0^*)}{k_{l2}} + \frac{k_{l1}}{k_{l2}^2} \cdot \log\left(\frac{k_{l1} + k_{l2} \cdot A_0^*}{k_{l1} + k_{l2} \cdot A}\right) = \frac{(B^* - B_0^*)}{k_{l4}} + \frac{k_{l3}}{k_{l4}^2} \cdot \log\left(\frac{k_{l3} + k_{l4} \cdot B_0^*}{k_{l3} + k_{l4} \cdot B}\right), \qquad (24)$$

where $A^*$ and $B^*$ are the total number of alive individuals of species $A$ and $B$ respectively, $A_0^*$ and $B_0^*$ are the initial conditions, and $k_{l1}$, $k_{l2}$, $k_{l3}$ and $k_{l4}$ are the parameters of the model. We note that, comparing Eqs (1) and (2) with Eq. (24), in the modified Lanchester model we have the sum of a linear term plus a logarithmic one. In Figures 12 and 13 we report the comparison, in terms of the total number of alive individuals of the species *A* and *B*, between the two deterministic models and the observations of the experimental results. The agreement of the chemical scheme compares with experimental data better than the modified Lanchester model (see Table 3), but not sufficiently to completely characterize the ant battle dynamics, *i.e.*, the latter model does not permit to use all experimental data as it consider only the population trend of the two species and not the groups that form during the battle.

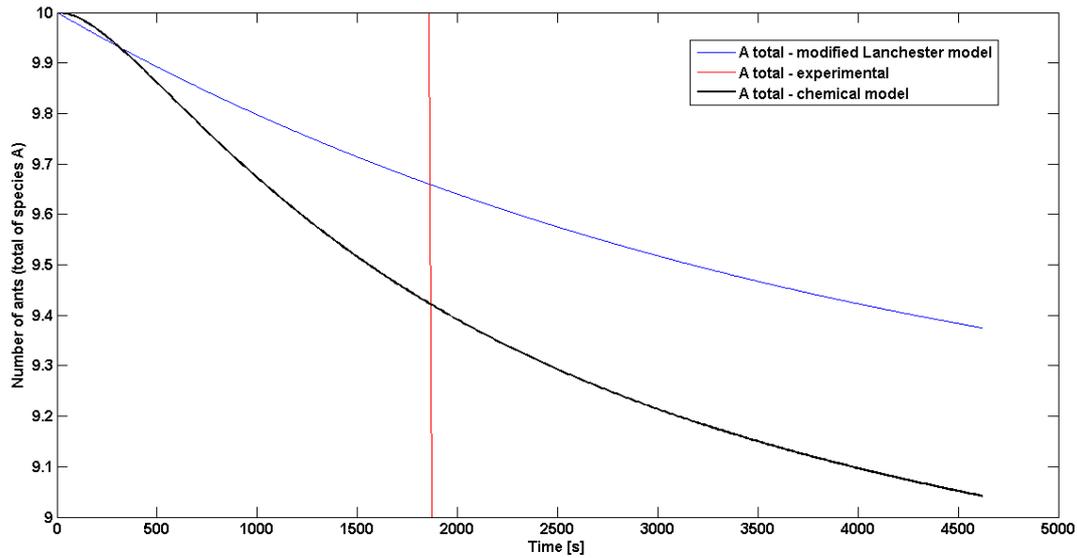

**Fig. 12.** The experimental data vs. the solution of the deterministic model and the Lanchester one both identified by means of SFA (Simplex Flexible Algorithm) for the chemical species A.

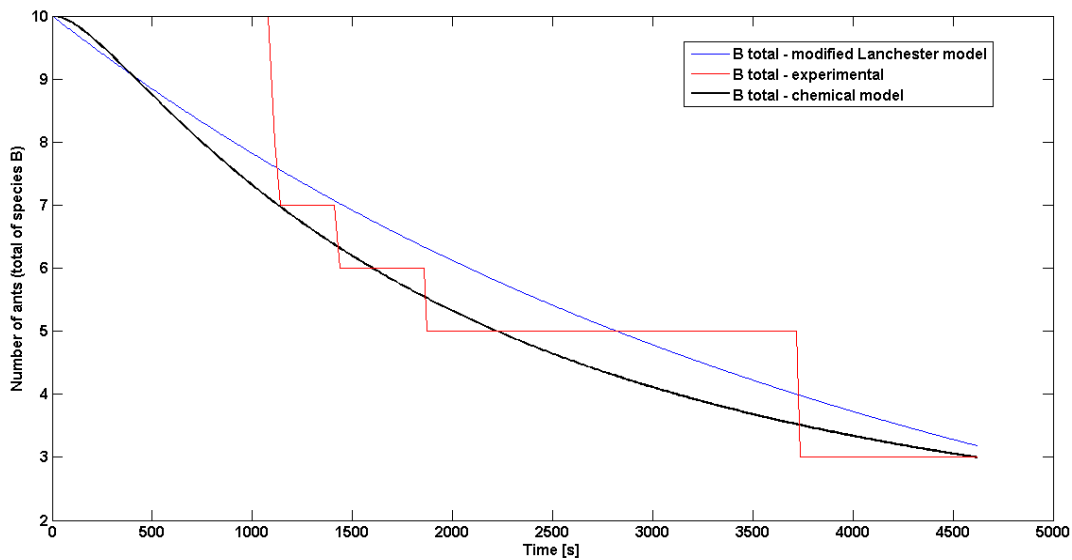

**Fig. 13.** The experimental data vs. the solution of the deterministic model and the Lanchester one both identified by means of SFA (Simplex Flexible Algorithm) for the chemical species B.

|  | Modified Lanchester vs. Experimental Data | Chemical Model vs. Experimental Data |
|---|---|---|
| Mean error species A | 0.3722 | 0.2353 |
| Mean error species B | 0.7217 | 0.7106 |

**Table 3.** A comparison between the performance of the Chemical Model and the modified Lanchester one.

We therefore adopted the stochastic procedure, as described in the sub-section III of the previous section. The distribution of 100 values obtained for each reaction constant $k_j$ are fitted by a Gaussian curve; we show in the Figure 14 some examples. The mean value of each parameter $k_i$, obtained by stochastic simulation, is reported in Table 2. These values are similar to the corresponding true values, *i.e.*, the values of the best parameters that were used to perform the stochastic simulations with the Gillespie algorithm. To perform a comparison between DE and the stochastic model (that produces data irregularly spaced in time), we first average the data from each simulation over equally spaced temporal bins $\Delta t$, and then perform the same-time averages over all the simulations. The resulting graphs are plotted in Figure 15 for the species A and B respectively, considering also the standard deviation and the solutions of DE. The small differences between the averages of the stochastic simulations and DE are due to finite size effects.

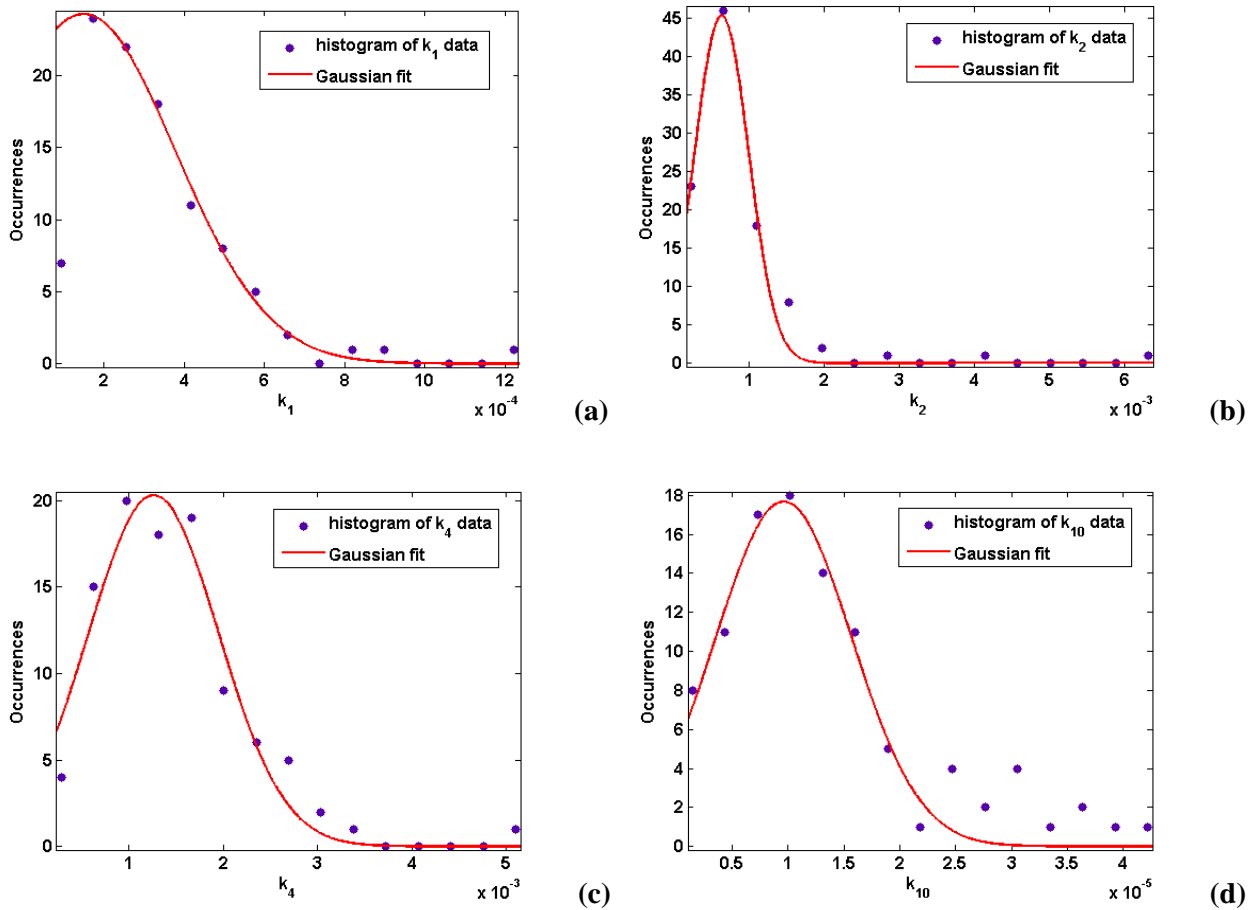

**Fig. 14.** The Gaussian distribution of the reaction coefficient $k_1$ (a), $k_2$ (b), $k_4$ (c), $k_{10}$ (d) with our stochastic procedure.

Moreover, the variability in this type of systems leads us to consider a stochastic approach to evaluate a probability of success of a species changing the initial condition. In Figure 16 we show, in the phase plane (total of A vs. total of B), an analysis to the initial conditions of the deterministic model simulating up to the extinction of a species. Starting from 10A vs. 10B up to 5A vs. 15B, and considering the calibrated parameters of Table 2 (column 1), we observe a zone, crossed by a dash-line and a dot-line (initial conditions with only mathematical sense [5A, 13.4B] and [5A, 13.8B] respectively) and between the curves with initial conditions [5A, 13B] and [5A, 14B], in which the fluctuations are dominant and the deterministic model is not sufficient to establish the winning species. Averaging over 1000 simulations we get, in case of 5A vs. 14B, that species A loses (dies) with 52.9 % of probability, while, in case of 5A vs. 13B with 39.4 % and in case of 5A vs. 15B with 66 %.

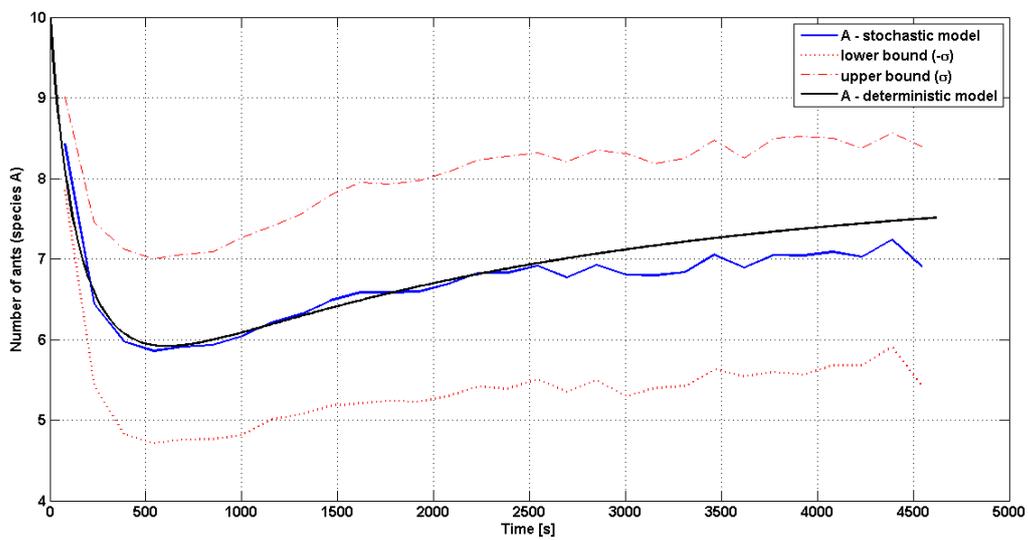

(a)

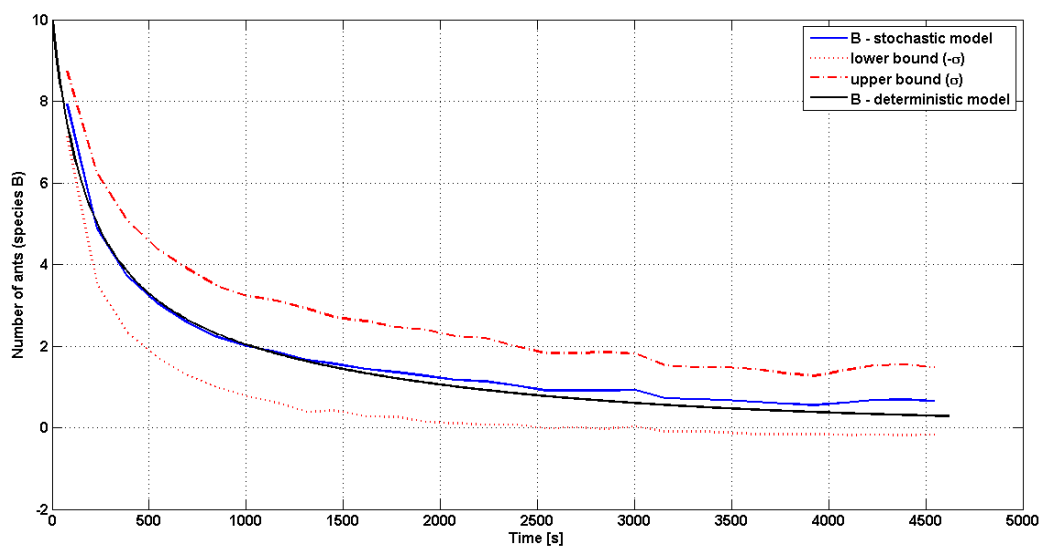

(b)

**Fig. 15.** The stochastic model obtained as average of 1000 simulations with Gillespie algorithm vs. deterministic one, considering also the variance, for the chemical species A (a) and B (b).

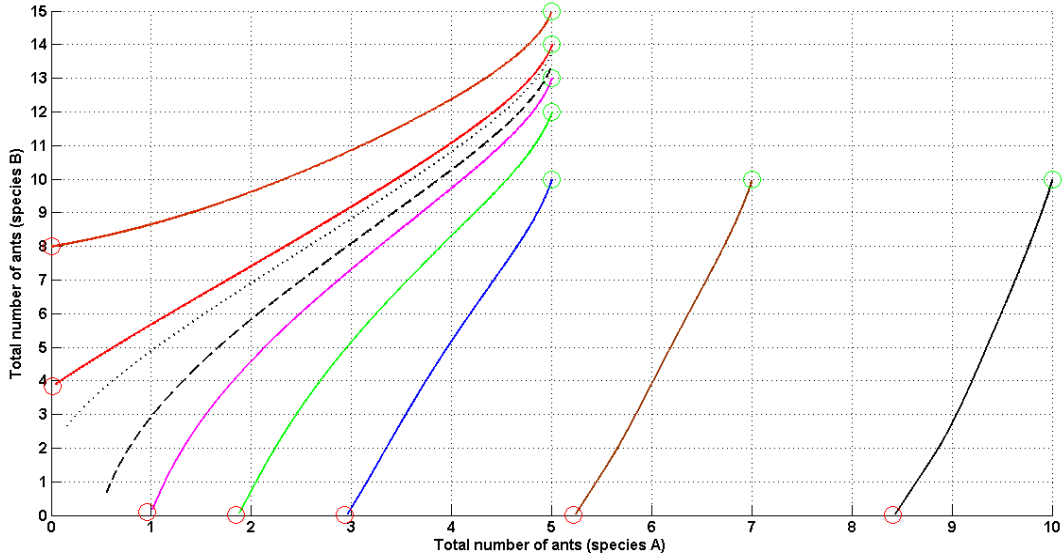

**Fig. 16.** Analysis of the initial condition in order to study the supremacy of a species over the other one.

## 5. Discussion and conclusions

We performed laboratory experiments to study the behavior or strategy of two species of ants during their fight and we propose a methodology to describe the battle dynamics and to estimate the probability of extinction of a species in case of conflict with the other one. We focused on 10 vs. 10 (10 *L. Paralienus*, 10 *L. Neglectus*) battles, but we studied, with less statistics, the case 5 vs. 15 (5 *L. Paralienus*, 15 *L. Neglectus*). As mentioned, *L. Neglectus*, despite the smaller size than *L. Paralienus*, is more aggressive and can attack in group. In order to extract quantitative data from our video observations, we develop a model based on chemical reactions (Eqs. 3-12) that take into account the observed dynamics, *i.e.*, the "strategy" of *L. Neglectus* ants that can attack the other species with up to three individuals, while *L. Paralienus* defends itself as a single individual. We considered the sufficiently long-lasting groups as chemical compounds. We therefore derived a system of non-linear chemical differential equations for the defined chemical species. Then, using the SFA and the data obtained from an experiment of 10 vs. 10 individuals, we achieve the parametric identification of the chemical model, *i.e.*, we find the set of best parameters for which the functional error (Eq. 18) is minimized. With the same procedure we identify the parameters for the modified Lanchester model, Eq. (23), that is used as benchmark to test the goodness of the chemical model, *i.e.*, the system represented by Eqs. (13)-(17), as shown in Figures 12 and 13. The chemical approach appears to be more precise than Lanchester (see also Table 3). Moreover the coefficients $k_i$, obtained by parametric calibration of the differential system, Eqs. (13)-(17), allow to draw some considerations. First of all *L. Paralienus* (species A), due to its size, has more possibilities to defeat *L. Neglectus* (species B) in a duel: indeed the parameter $k_4$ of reaction AB→A is larger ($1.2 \cdot 10^{-3}$) than all other parameters, while the death of B is less probable, than the latter reaction, considering ABB→AB regulated by parameter $k_7$ ($4.5 \cdot 10^{-5}$). On the contrary, the death of A is more probable, than AB→B ($k_3 = 7.7 \cdot 10^{-5}$), considering the reaction ABBB→3B, where $k_{13} = 0.9 \cdot 10^{-3}$ is comparable with $k_4$. Therefore, considering the coefficients of reactions that bring to death, we observe that the strategy of *L. Neglectus* in 10 vs. 10 is not sufficient to defeat *L. Paralienus*. The same consideration is deduced with the stochastic model performing 1000 simulations with the Gillespie algorithm with the optimized parameters, in which we observe zero successes of the species B, *i.e.*, all B individuals die, while species A survives with a certain mortality. Moreover, the analysis in the phase plane of the deterministic model, varying the initial condition, shows the importance of the initial ratio between the opponents. Then the stochastic approach is fundamental to estimate the fluctuations, in term of individual number belonging to chemical species, that can allow the defeat of a species. Finally, the preliminary experiments described served for the development of the stochastic based-Gillespie model. We are actually carrying on more detailed experiments, possibly analyzed using automatic tracking systems and involving more ant species.

# References

Batchelor T.P. and Briffa M., 2010. Influences on resource-holding potential during dangerous group contests between wood ants, Animal Behaviour, 80, 443-449.

Batchelor T.P., Santini G., Briffa M., 2011. Size distribution and battles in wood ants: group resource-holding potential is the sum of the individual parts. Animal Behaviour 83, 111-117.

Bracken J., 1995. 'Lanchester Models of the Ardennes Campaign'. Naval Research Logistics, 42, 559-577.

Campillo F. and Lobry C., 2012. Effect of population size in a predator–prey model. Ecological Modelling. 246, 1– 10.

Cash J.R. and Karp A.H., 1990. "A variable order Runge-Kutta method for initial value problems with rapidly varying right-hand sides", ACM Transactions on Mathematical Software 16: 201-222.

Cremer S., Ugelvig L.V., Drijfhout F. P., Schlick-Steiner B.C., Steiner F.M., Seifert B., Hughes D.P., Schulz A., Petersen K.S., Konrad H., Stauffer C., Kiran K., Espadaler X., d'Ettorre P., Aktac N., Eilenberg J., Jones G.R., Nash D.R., Pedersen J.S., Boomsma J.J., 2008. The evolution of invasiveness in garden ants. Plos one 3(12).

Franks N.R. and Partridge L.W., 1992. Lanchester battles and the evolution of combat in ants. Animal Behavior., 45, 197-199.

Friker R.D.Jr., 1998. Attrition Model of the Ardennes Campaign. Naval Research logistic Vol. 45.

Gillespie D.T., 1977. Exact Stochastic Simulation of Coupled Chemical Reaction. The Journal of Phisical Chemistry, Vol. 81, No.25.

Johnson I.R., MacKay N.J., 2011. Lanchester Models and the Battle of Britain. *Naval Research Logistics*, Vol. 58.

Lanchester F. Aircraft in Warfare, Apleton. New York 1916

Marsili-Libelli S., 1992. Parameter estimation of ecological models, Ecol. Model., 62, 233–258.

McGlynn T.P., 2000. Do Lanchester 's laws of combat describe competition in ants? Behavioral Ecology Vol.11 No 6: 686-690.

Plowes N.J.R. and Adams E.S., 2005. An empirical test of Lanchester's square law: mortality during battles of the fire ant *Solenopsis invicta.* Proc. R. Soc. B 272, 1809 – 1814.

Seifert B., 1992. A taxonomic Revision of the Palaearctic Members of the ant subgenus Lasius s. tr.(Hymenoptera, Formicidae).

Seifert B., 2000. Rapid range expansion in Lasius Neglectus (Hymenoptera, Formicidae)- an Asian invader swamps Europe. Mitt.Mus. Nat.kd. Berl. Dtsch.entomol. Z.47, 173-179.

Tsutsui N.D. and Suarez A.V., 2003. The Colony Structure and Population Biology of
Invasive Ants  Conservation Biology, Pages 48–58 Vol 17, No.1.
Ugelvig L.V., Falko P., Drijfhout F.P.,Daniel JC, Kronauer D. JC, Jacobus J., Boomsma J.J., Pedersen J.S., and Cremer S., 2008. The introduction history of invasive garden ants in Europe: Integrating genetic, chemical and behavioural approaches. *BMC Biology* 2008, **6**:11.

Whitehouse M.E.A  & Jaffe  K. 1996. Ant wars: combat strategies, territory and nest defence in the leaf-cutting ant *Atta laevigata. Anim. Behav.*, 51**,** 1207–1217.

Wilson M.L., Britton N.F. and Franks N.R., 2001. Chimpanzees and the mathematics of battle. *Proc. R. Soc. Lond. B* 2002 269, 1107-1112.